\documentclass[aps,prl,cha,graphicx,amsmath,amssymb,reprint,twocolumn,
superscriptaddress,floatfix]{revtex4}
\usepackage[]{graphicx}

\begin{document}
\title{Actively crosslinked microtubule networks: mechanics, dynamics and
filament sliding}
\author{Sebastian F\"urthauer}
\affiliation{Center for Computational Biology, Flatiron Institute, New York, NY 10010, USA} 
\author{Bezia Lemma}
\affiliation{Department of Physics, Harvard University, Cambridge, MA 02138, USA}
\affiliation{Department of Physics, Brandeis University, Waltham, MA 02454, USA}
\affiliation{Department of Physics, University of California, Santa Barbara, CA 93106, USA}
\author{Peter J. Foster}
\affiliation{Physics of Livings Systems, Department of Physics, Massachusetts
Institute of Technology, Cambridge, MA 02139, USA}
\author{Stephanie C. Ems-McClung}
\affiliation{Medical Sciences Program, Indiana University, Bloomington, IN 47405, USA}
\author{Claire E. Walczak}
\affiliation{Medical Sciences Program, Indiana University, Bloomington, IN 47405, USA}
\author{Zvonimir Dogic}
\affiliation{Department of Physics, Brandeis University, Waltham, MA 02454, USA}
\affiliation{Department of Physics, University of California, Santa Barbara, CA 93106, USA}
\author{Daniel J. Needleman}
\affiliation{Paulson School of Engineering \& Applied Science and Department of Molecular \& Cellular Biology, Harvard
University, Cambridge, MA 02138, USA}
\author{Michael J. Shelley}
\affiliation{Center for Computational Biology, Flatiron Institute, New York, NY 10010, USA}
\affiliation{Courant Institute, New York University, New York, NY 10012, USA}
\begin{abstract}
    \noindent 
Cytoskeletal networks are foundational examples of active matter and central to
self-organized structures in the cell. {\it In vivo}, these networks are active
and heavily crosslinked. Relating their large-scale dynamics to properties of
their constituents remains an unsolved problem.  Here we study an {\it in vitro}
system made from microtubules and XCTK2 kinesin motors, which forms an aligned
and active gel. Using photobleaching we demonstrate that the gel's aligned
microtubules, driven by motors, continually slide past each other at a speed
independent of the local polarity. This phenomenon is also observed, and remains
unexplained, in spindles. We derive a general framework for coarse graining
microtubule gels crosslinked by molecular motors from microscopic
considerations. Using the microtubule-microtubule coupling, and force-velocity
relationship for kinesin, this theory naturally explains the experimental
results: motors generate an active strain-rate in regions of changing polarity,
which allows microtubules of opposite polarities to slide past each other
without stressing the material.
\end{abstract}
\maketitle

Active materials are made from energy consuming constituents whose
activity keeps the system out of equilibrium. Their study has deepened
our understanding of self-organizing processes, both {\it in vivo} and
{\it in vitro} \cite{marchetti2013hydrodynamics, needleman2017active}.  
Its foundational examples include suspensions of
microswimmers, {\it in vitro} assemblies of purified cellular components, and
the cell cytoskeleton \cite{alberts_2002_book}.  While notable progress has been
made in deciphering the non-equilibrium physics of active materials by using
experiments and symmetry-based phenomenological theories \cite{kruse2005generic,
joanny2007hydrodynamic, julicher2018hydrodynamic}, there is far less
understanding of how the large-scale dynamics actually devolves from microscopic
activity. Having such theoretical frameworks would be powerful, allowing the
development of unified analysis and design tools for new active materials, and
understanding the biology of the cell cytoskeleton. Here we address this
challenge for systems made of cytoskeletal polymers which are fully percolated
by moving molecular motors. First, we report
observations of a dense nematically-aligned active gel made from
purified microtubules and XCTK2 kinesin motors. Driven by these
motors, its microtubules continually slide within the gel. Using photo
bleaching we show the surprising result that their sliding speed is
independent of the local gel polarity. This independence of speed from
polarity is also observed in {\it Xenopus} meiotic spindles
\cite{mitchison2005mechanism,burbank2007slide,yang2008regional} and is
an unexplained feature that suggests an robust internal coupling of
the microtubular material. To investigate this phenomenon, we
introduce a framework for deriving continuum theories for heavily
crosslinked, active gels from microscopic considerations. We use this
framework to obtain a theory for the XCTK2-microtubule system which
explains our experimental findings without adjustable parameters.

How large-scale behaviors of actively crosslinked networks emerge from
the properties of their constituents is an actively researched
topic in biology \cite{naganathan2018morphogenetic,
  roostalu2018determinants}. Many aspects of cell biology, including
cell shape, motility, and division, are driven by the cytoskeleton
\cite{bray2001cell}. The cytoskeleton consists of polar filaments -
mainly actin and microtubules - and the proteins which crosslink them
and organize their behavior. Molecular motors are active crosslinkers
that use chemical energy to move filaments relative to each other.
They play a central role in determining the architecture and dynamics
of cytoskeletal structures such as the cell cortex and the spindle
\cite{alberts_2002_book}. The XCTK2-microtubule system we present here
recapitulates the previously unexplained polarity independent sliding
motion of microtubules that was observed using speckle microscopy in
{\it Xenopus} meiotic spindles
\cite{mitchison2005mechanism,burbank2007slide,yang2008regional} and
allows us to study how polarity independent filament motion can emerge
in actively crosslinked networks.

Phenomenological theories for actively crosslinked networks
\cite{kruse2005generic, joanny2007hydrodynamic, furthauer2012active,
    marchetti2013hydrodynamics,julicher2018hydrodynamic} have been
derived from conservation laws and symmetry considerations. They can
reproduce the pattern of flows observed in artificial \cite{thampi2013velocity} and
biological systems such as the cell cortex
\cite{salbreux2009hydrodynamics, mayer2010anisotropies,
  naganathan2014active,naganathan2018morphogenetic} and quantitatively
explain some aspects of spindle structure and dynamics
\cite{brugues2014physical}. However, such theories do not address how
the large-scale behaviors of actively crosslinked networks emerge from
the properties of its constituents. For this, an approach that derives
macroscopic material laws from the properties of motors and filaments
is needed.

Previous efforts to derive continuum theories for active gels from microscopic
interactions \cite{kruse2000actively,aranson2005pattern,liverpool2005bridging,
    liverpool2008hydrodynamics, saintillan2008instabilitiesPRL,
    saintillan2008instabilitiesrPHYSFLUID, saintillan2013active,
foster2015active, gao2015multiscalepolar,heidenreich2016hydrodynamic,
maryshev2018kinetic} have considered dilute systems, in which individual
motor-filament clusters are thought of as disconnected objects. In contrast,
actively crosslinked networks, such as the cytoskeleton\cite{alberts_2002_book,
belmonte2017theory} are crosslinked over scales comparable to the system size
and display behaviors different from those predicted by a dilute theory. 

\begin{figure}[h] \centering
    \includegraphics[width=0.8\columnwidth]{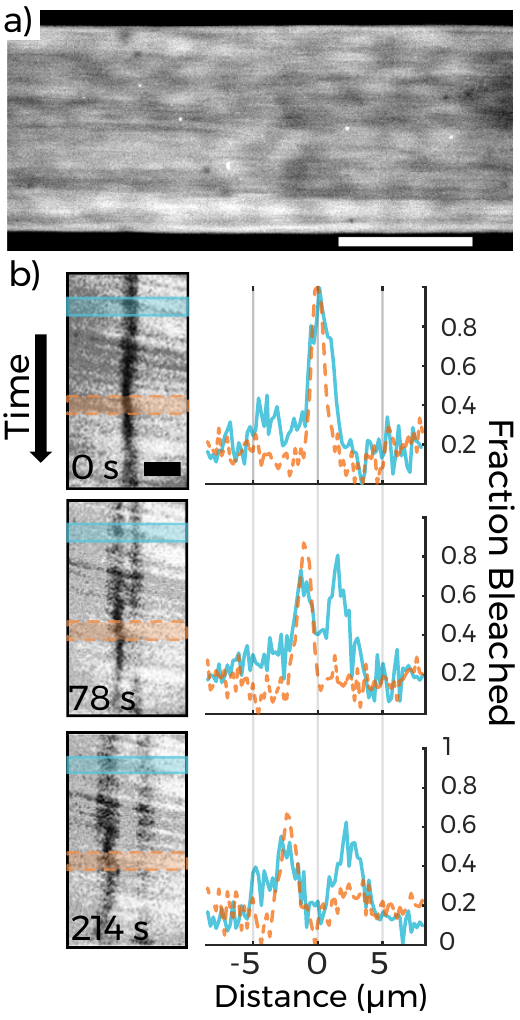} \caption{
      Bleaching experiments on an aligned gel of stabilized
      microtubules and XCTK2 molecular motors.  (a): An image
      of the aligned gel within a microfluidic chamber.  $50\mu\mathrm{m}$ scale
      bar. (b): (left) Higher magnification image of a bleached
      line within the above material. $5\mu\mathrm{m}$ scale bar. The bleached line splits
      into two parallel lines as the microtubules slide apart. (right)
      Line scans of the bleached image highlighting a region of high
      polarity (orange, dotted line, $P_{exp} =0.63\pm0.12$ ) and a neighboring region of low
      polarity (blue, solid line, $P_{exp} = 0.08\pm0.17$ ) While the polarity of these two
      regions greatly differs, the peaks move apart at nearly
      identical speeds.  } \label{fig:experiment} \end{figure}
\begin{figure}[h] \centering
    \includegraphics[width=0.99\columnwidth]{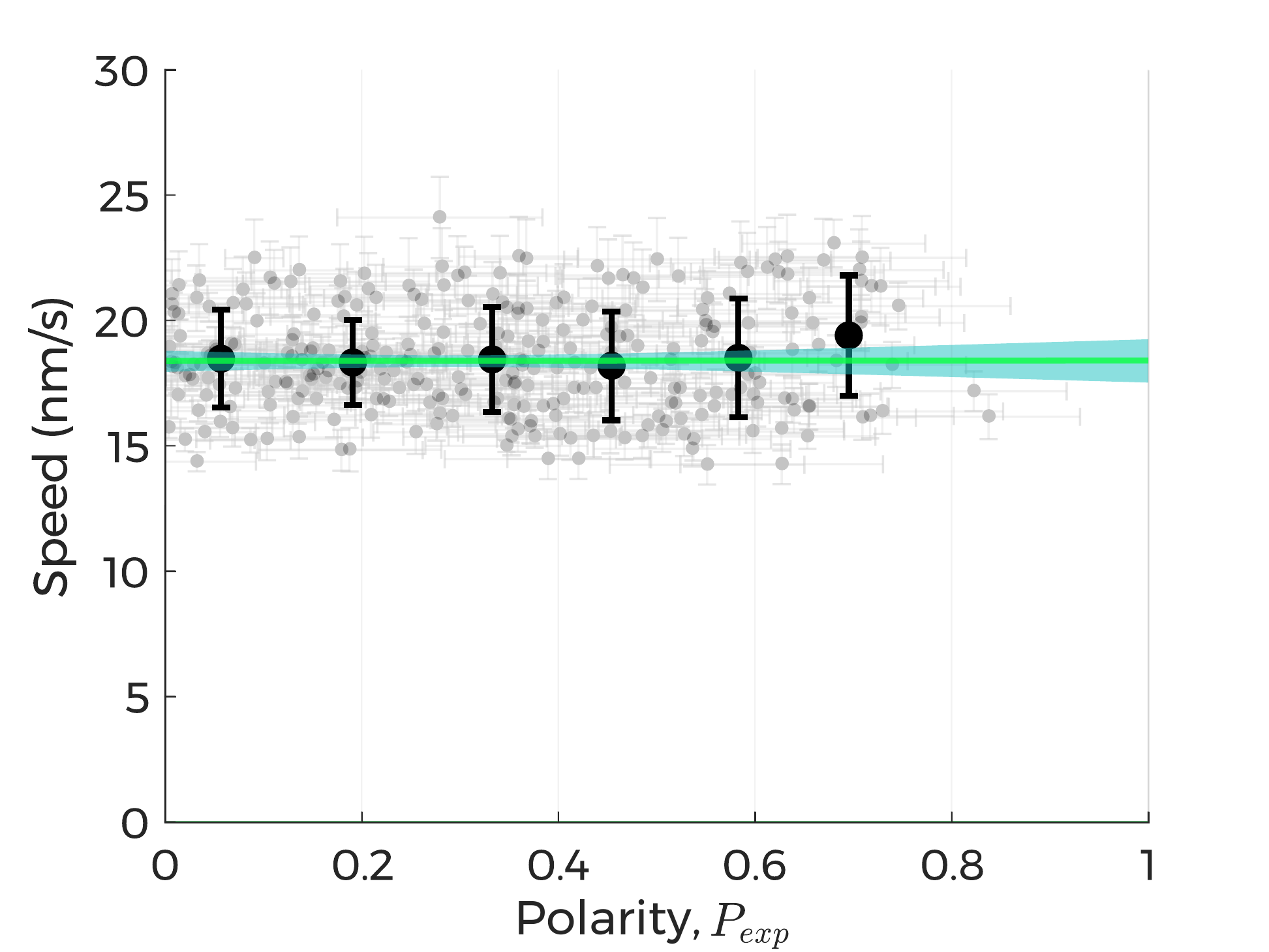}
    \caption{ Sliding speed {\it vs.} polarity from photobleaching
      experiments. Each gray dot corresponds to a local measurement of
      sliding speed and polarity from a bleach line, as in Fig. 1
      ($n=413$). Error bars in the individual velocities correspond to $95\%$
      confidence intervals of a linear regression of positions of the bleached
      line over time.  Error bars in the individual polarities correspond to the
      standard deviation of the local polarity over time. Large black dots show
      measurements binned by polarity with error bars showing the corresponding
      standard deviation. In solid green is a regression line, and shaded behind
      it is the 95\% confidence interval of predicted responses to the
      regression. The Pearson correlation coefficient is 0.08 with the $95\%$
      confidence bounds between $-0.08$ and $0.15$ (not significantly different
      from $0$, p-value$ = 0.51$).
    } \label{fig:experiment2}
\end{figure} 

\noindent{\bf Experiments:} To study the behavior of heavily
crosslinked active gels we created one using a {\it in vitro} system
of purified components. We made solutions of fluorescently labeled
tubulin and fluorescently labeled XCTK2, a Kinesin-14 molecular motor
capable of crosslinking and sliding aligned microtubules
\cite{hentrich2010microtubule}.  We added paclitaxel to the solution,
which nucleates and stabilizes microtubules, and rapidly loaded the
sample into a rectangular microfluidic chamber (see Methods for details).

The microtubule-motor mixture spontaneously self-organizes into a macroscopic
gel, in which microtubules are nematically aligned parallel to the long axis of
the microfluidic chamber (Fig.~\ref{fig:experiment}A). Once the microtubules are
aligned they display continuous polar motion, sliding along the materials
director axis. The activity of the gel results in a buckling instability on time
scales much longer than those considered here.     

We observe this material motion parallel to the director by photo-bleaching the
microtubules. As bleaching marks only a subset of microtubules, it allows their
subsequent motions to be monitored. We used a femto-second Ti:Sapphire laser to
bleach lines orthogonal to the direction of nematic alignment. Control
experiments confirmed that the laser bleached, but did not ablate, the
microtubules (see Methods). The bleached material splits into two parallel lines
which move apart along the direction of alignment
(Fig.~\ref{fig:experiment}B,left), indicating that microtubules in the gel are
continually sliding relative to each other.  

When a bleached line splits into two, the relative fluorescent intensity of the
two new lines reflects the relative number of left-moving and right-moving
microtubules. Hence the relative bleach intensity provides a measure of the
polarity at the location of the bleach.  We define experimental polarity as
$P_{exp} =|\frac{ A_1 - A_2}{A_1 + A_2}|$, where $A_1$ and $A_2$ are the
amplitudes of bleach intensities of the two lines. The speed at which the two
new lines move apart provides a measure of the speed of microtubule sliding.
Along the same bleach line, some regions have very high polarity
(Fig.~\ref{fig:experiment}B, orange, $P_{exp}=0.63\pm0.12$), while others have
very low polarity (Fig.~\ref{fig:experiment}B, blue, $P_{exp}=0.08\pm 0.17$).

Despite differences in polarity, the speed of bleach line remains fairly
constant, (see Fig.~\ref{fig:experiment2}). 
This result argues that the microtubule sliding speed has little to no
dependence on polarity in these gels.  This is similar to the independence of
microtubule sliding speed and polarity observed in spindles
\cite{mitchison2005mechanism,burbank2007slide,yang2008regional}, where they have
been quantified using speckle microscopy.  Bleach lines in these gels with XCTK2
move apart at an average speed of $18.6\pm0.9\mathrm{nm/s}$, very close to the
observed speed of $\sim 20\mathrm{nm/s}$ that XCTK2 slides apart isolated pairs
of microtubules \cite{hentrich2010microtubule}.

\begin{figure}[h]
    \centering
    \includegraphics[width=0.9\columnwidth]{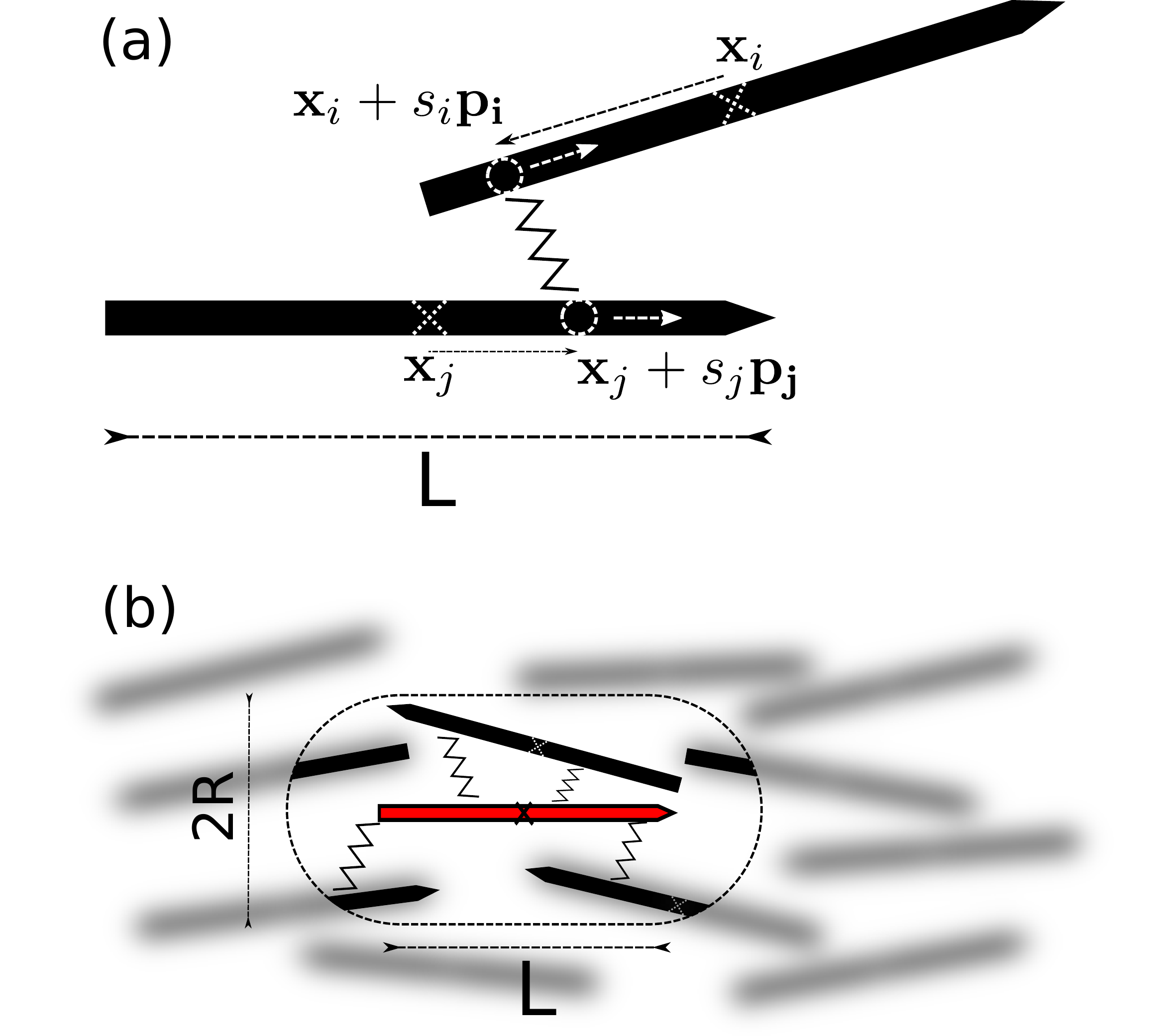}
    \caption{Sketch of the microscopic model: (a) each microtubule
      (black solid arrow) is characterized by its center of mass
      position $\mathbf x$ (white crosses) and its orientation
      $\mathbf p$. Microtubules interact via motor molecules (zig-zag
      line) which connect them at positions $s_i, s_j$ (white circles)
      on and walk towards their plus ends (white arrows). \
      (b) In the gel crosslinkers of a test-microtubule (red)
      explore all possible connections to other microtubules within
      reach (within the dashed capsule), i.e within a sphere of radius
      $R$ around any position along the test-microtubule.}
    \label{fig:theory}
\end{figure}

The polarity independent sliding speed of microtubules is difficult to reconcile
with current theories of cytoskeletal filaments and molecular motors. These theories
apply in the case of dilute and sparsely crosslinked networks, in which
the length scale over which filaments are crosslinked is small compared to the
system size. To elucidate the issue, we follow the arguments presented in
\cite{kruse2000actively, gao2015multiscalepolar} which consider a dilute
collection of sharply aligned microtubules interacting via molecular motors. In
the absence of external driving forces, balance of forces yields
$
  \rho^+ f^+ + \rho^- f^- = 0,
$
where $\rho^\pm$ are the densities of microtubules pointing along the
nematic axis of the system in the positive and negative directions,
respectively, and $f^\pm$ are the forces acting on microtubules from their
interactions with $\mp$ microtubules, respectively. In a dilute system the force
exerted by a motor on a microtubule is balanced only by the drag between the
microtubule and the surrounding medium and, assuming the medium is locally at
rest, the velocity on microtubules $v^\pm$ will be given by $v^\pm = \mu f^\pm$,
where $\mu$ is the microtubule-mobility. If the molecular motor crosslinking the
microtubules moves at a speed $V$, this imposes $v^+-v^- = 2V$. Taken together,
this leads to $v^\pm=\mp V(1\mp P)$, where
$P=\frac{\rho^+-\rho^-}{\rho^++\rho^-}$ is the polarity. 

That is, the sliding velocity of microtubules is predicted to vary linearly with
polarity, in conflict with what is observed in our experiments as well as in
spindles. Fundamentally, this result is a consequence of force balance in dilute
systems: the mass fluxes of left moving and right moving microtubules locally
balance. This generalizes to more sophisticated versions of dilute theories that
include additional effects, such as long-range hydrodynamics. The disagreement
between this theory and experiment suggests that this system may not be in a
dilute regime. In order to probe this possibility, we quantified the microtubule
density in our gel by measuring the initial molarity of components in our
mixture and then determined the percentage of those components which were
incorporated into the final gel via fluorescence microscopy. We estimate a 5\%
volume fraction of polymerized microtubules in the gels, with 17 microtubules
per $\mu$m$^3$ (See Methods) and $\sim 25$ XCTK2 dimers bound to each microtubule.
Since microtubules are significantly longer than their average spacing, this
result argues that these networks are dense and heavily crosslinked by molecular
motors--far from the dilute theory described above. Thus, we introduce a theory
of heavily crosslinked networks of microtubules and molecular motors.

\noindent{\bf Theory for a heavily crosslinked microtubule gel.}  
In our theory, microtubules are characterized by their length $L$, and their
velocities $\mathbf v_i$ and orientations $\mathbf p_i$.
Generically the distribution of microtubules $\psi(\mathbf x, \mathbf p) =
\sum\limits_i\delta(\mathbf x - \mathbf x_i)\delta(\mathbf p-\mathbf p_i)$ obeys
the Smoluchowski equation $\partial_t\psi = -\mathbf\nabla \left( \dot{\mathbf
x}\psi\right) -\partial_{\mathbf p}\left( \dot{\mathbf p} \psi \right).$ The
microtubules translational $\dot{\mathbf{x}}\left(\mathbf x_i, \mathbf
p_i\right) =\mathbf v_i$ and rotational $\dot{\mathbf{p}}\left( \mathbf x_i,
\mathbf p_i\right) = \dot{\mathbf p}_i$ fluxes can be determined from the force
balance $\mathbf F_i = 0$ and torque balance $\mathbf T_i =0$ conditions which
state that the total force $\mathbf F_i$ and torque $\mathbf T_i$ on each
microtubule vanish. 

The forces and torques applied to microtubules are generated by molecular motors
which actively crosslink them. Here, motors are characterized by their
force-velocity relation, the distance $R$ (the motor size) over which they can
crosslink filaments, and the crosslinking torque that they apply. The force that
microtubule $j$ exerts on microtubule $i$, via active crosslinks between
$\mathbf x_i + s_i \mathbf p_i$ and $\mathbf x_j + s_j \mathbf p_j$ (see
Fig.~\ref{fig:theory}a), is given by $\mathbf{F}_{ij}$ and the total force on
microtubule $i$ is
\begin{eqnarray}
  \mathbf{F}_i = \int\limits_{-L/2}^{L/2} ds_i
  \sum\limits_j
    \int\limits_{-L/2}^{L/2} ds_j 
    \int d\mathbf p_j
    \int\limits_{\Omega(\mathbf x_i + s_i \mathbf p_i )}dx^3
    \nonumber\\
    \times
    \delta(\mathbf x -\mathbf x_j-s_j \mathbf p_j)\delta(\mathbf p -\mathbf p_j)
    \mathbf{F}_{ij},
    \label{eq:MTForce}
\end{eqnarray}
where $\Omega(\mathbf x)$ denotes a sphere of radius $R$ centered at $\mathbf x$
(see Fig.~\ref{fig:theory}b). We note that the force density can be written as
the divergence of the network stress tensor $\mathbf{\Sigma}$ according to
Kirkwood's formula since $\mathbf{F}_{ij}=-\mathbf{F}_{ji}$. Thus, the force
balance of the network reads
\begin{equation}
    \mathbf\nabla\cdot \mathbf{\Sigma}(\mathbf x) = \sum\limits_i \mathbf F_i \delta(\mathbf
    x-\mathbf x_i) = \mathbf 0.
    \label{eq:force_density}
\end{equation}
The torque that a crosslinker between filaments $i$ and $j$ exerts on filament
$i$ is given by $\mathbf T_{ij} = s_i\mathbf p_i\times \mathbf F_{ij} +
\mathbf\Gamma_{ij}$, a where $\mathbf\Gamma_{ij}$ is the contribution stemming
from an explicit crosslinker torque. The total torque $\mathbf T_i$ on
microtubule $i$ obeys an equation analogous to Eq.~{(\ref{eq:MTForce})}, (see
Methods).

We model the XCTK2 motors by a force-velocity relation
$\mathbf{F}_{ij}(\Delta\mathbf{v}_{ij})$, which depends on the velocity
difference $\Delta\mathbf{v}_{ij}$ between the attachment points of the motor on
the two microtubules, and a crosslinking torque $\mathbf{\Gamma}_{ij}(\mathbf
p_i, \mathbf p_j)$ which depends on their orientations. For simplicity we ignore
the effects of non-uniform distributions of motors along microtubules. To linear
order in $\Delta \mathbf{v}_{ij}$,
\begin{eqnarray}
  \mathbf{F}_{ij} 
  &=& -\mathbf{G} \cdot
    \left[\mathbf v_i + s_i\dot{\mathbf p}_i + V\mathbf p_i
    - (\mathbf v_j + s_j\dot{\mathbf p}_j + V\mathbf p_j)
    \right],
\label{eq:force_velocity_relation}
\end{eqnarray}
where $V\mathbf p_i$ is the velocity of the motor relative to the microtubule it
is bound to. This choice renders the force between two microtubules dependent on
their relative polarities. The linear response coefficient $\mathbf{G}$ is the
motor friction and is in general a second rank tensor. In the following we
choose $\mathbf{G} = \gamma\mathbf{I}$ for simplicity. For the crosslinking
torque we choose $\mathbf \Gamma_{ij} = \nu\mathbf p_i \times\mathbf p_j
(\mathbf p_i\cdot \mathbf p_j)$, which aligns microtubules, as would for
instance a torsional stiffness of the crosslinker. The coefficient $\nu$
characterizes the magnitude of this effect. It is straightforward to generalize
our formulation to include motor activity which varies along the length of a
filament, say by pausing at microtubule ends as has been previously argued to
drive network contractions \cite{kruse2000actively, foster2015active,
foster2017connecting}.

We next derive a continuum theory for our system. We start by expanding
Eq.~{(\ref{eq:MTForce})} around the center-of-mass positions $\mathbf{x}_i$ and
$\mathbf{x}_j$, 
\begin{eqnarray}
  \mathbf F_i &=& - \gamma L^2 \int\limits_{\Omega(\mathbf x_i)} dx^3
  \rho(\mathbf x)
  \left[(\mathbf v_i -\mathbf v(\mathbf x))
  + V(\mathbf p_i-\mathbf P(\mathbf x))\right]
  \nonumber\\
  &-& \gamma \frac{L^4}{12} \int\limits_{\Omega(\mathbf x_i)} dx^3
  \mathbf\nabla\cdot\left[\rho(\mathbf x)(\mathbf p_i\dot{\mathbf p}_i
  +\mathcal{H}(\mathbf x))\right] 
    \label{eq:MTiforce1}
\end{eqnarray}
Here we assumed that the length-scale $\ell$ of gradients in the
system is large compared to the microtubule length $L$ and dropped
terms with more than one spatial derivative (see Methods). We
further introduced the density
\begin{equation}
    \rho(\mathbf x) = \int d\mathbf p\sum\limits_i\delta(\mathbf x-\mathbf x_i)\delta(\mathbf
    p- \mathbf p_i),
    \label{eq:density_definition}
\end{equation}
the polarity $\mathbf P(\mathbf x) = \left<\mathbf p_i\right>$,

the rotation rate tensor $\mathcal H(\mathbf x) = \left<\mathbf p_i \dot{\mathbf
p}_i\right>$ and the velocity field $\mathbf v(\mathbf x) = \left<\mathbf
v_i\right>$, where the
angled brackets
$\left<\ldots\right> = 1/\rho\int d\mathbf p\sum\limits_i\delta(\mathbf x-\mathbf x_i)\delta(\mathbf
p- \mathbf p_i) \ldots$ denote local averaging.

The force balance Eq.~{(\ref{eq:force_density})} becomes
\begin{eqnarray}
    &&\mathbf\nabla\cdot\mathbf{\Sigma}(\mathbf x) = \nonumber\\
    &-&\gamma L^2 \int\limits_{\Omega(\mathbf x)}
    dy^3\rho(\mathbf x)\rho(\mathbf y)\left\{\mathbf v(\mathbf y) -\mathbf v(\mathbf x) +V\left(\mathbf
    P(\mathbf y) -\mathbf P(\mathbf x)\right)\right\}\nonumber\\
    &-&\gamma \frac{L^4}{12} \int\limits_{\Omega(\mathbf x)}
    dy^3\rho(\mathbf x)\left(\mathbf\nabla\cdot\left(\rho(\mathbf y) \mathcal{H}(\mathbf y)\right) +
    (\mathbf\nabla\rho(\mathbf y))\cdot \mathcal{H}(\mathbf x)\right).
    \label{eq:force_density2}
\end{eqnarray}
Now, using that $R/\ell \ll 1$ we expand the integrand in
Eq.~(\ref{eq:force_density2}) around $\mathbf x$ and perform the integration. In
this way we obtain for the gel stress
\begin{equation}
    \mathbf{\Sigma} = \eta\rho^2(\mathbf \nabla\mathbf v + V\mathbf\nabla\mathbf P) -
    \alpha\rho^2 \mathcal{H}.
    \label{eq:derived_stress}
\end{equation}
The coefficients $\eta =\gamma\frac{4\pi}{15} L^2 R^5 $ and
$\alpha=\gamma\frac{4\pi}{36}L^4 R^3$ have dimensions of viscosity by density
squared.
The first term of the stress tensor in Eq.~{(\ref{eq:derived_stress})} is
viscous-like (i.e. depends on $\nabla\mathbf{v}$) and captures long
ranged coupling between microtubules. In contrast to a dilute
suspension theory, in which a viscous coupling would be induced by the
fluid, here it is induced by the crosslinkers. For active crosslinkers
($V\ne 0$) the stress-free state is self-straining and the gel's
spontaneous strain rate is $\mathbf\nabla\mathbf v =
-V\mathbf\nabla\mathbf P$. The second term of Eq.~{(\ref{eq:derived_stress})} is
generated by microtubules reorienting in the gel and is analogous to a nematic
alignment stress in a liquid crystal theory~\footnote{We note the ordering
    stress recovers the form derived for kinetic theories that approximate
    microtubule alignment by the Maier-Saupe free-energy
\cite{saintillan2013active} when using Eq.~{(\ref{eq:rotation_rates})}.}. 
  From Eq.~{(\ref{eq:MTiforce1})}, using
$\mathbf\nabla\cdot\mathbf{\Sigma}=0$, we obtain the flux balance of
microtubules 
\begin{equation}
    \mathbf v_i -\mathbf v = -V(\mathbf p_i -\mathbf P) -
    \frac{L^2}{12}\frac{\mathbf\nabla\rho}{\rho}\cdot(\mathbf p_i\dot{\mathbf{
    p}}_i -\mathcal{H}).
    \label{eq:final_flux}
\end{equation}

An analogous calculation for the torque balance (see Methods) yields
\begin{equation}
    \alpha \mathbf p_i \times \dot{\mathbf p}_i = \hat\nu \mathbf p_i\times \left(\mathbf p_i\cdot
    \mathcal{Q}\right),
    \label{eq:rotation_rates}
\end{equation}
which is reminiscent of Maier-Saupe theory and where $\hat\nu = \nu
L^2\frac{4\pi R^3}{3}$ and $\mathcal Q(\mathbf x) = \left<\mathbf p_i  \mathbf
p_i\right>$
is the nematic tensor order parameter. The force balance
$\nabla\cdot\mathbf{\Sigma} = \mathbf{0}$ with
Eqs.~(\ref{eq:derived_stress},\ref{eq:final_flux},\ref{eq:rotation_rates}),
fully specify the system's dynamics and can be used to time evolve the
distribution of microtubule positions and orientations.

\noindent{\bf Comparison and interpretation.} 
We next use this theory to investigate microtubule sliding in actively
crosslinked networks. Consider a fully aligned gel with all $\mathbf{p}_i =
\mathbf{\pm \hat e}$, for some unit orientation vector $\mathbf{\hat e}$. With
this, Eq.~({\ref{eq:rotation_rates}}) yields $\dot{\mathbf p}_i = 0$.  Thus, the
force balance $\nabla\cdot\mathbf{\Sigma} = \mathbf{0}$ and the flux balance Eq.
(\ref{eq:final_flux}) are solved by $\mathbf v_i = -V\mathbf p_i$, when applying
no stress boundary conditions at infinity. This means that all microtubules
translate with the motor velocity in the direction of their minus end. The
velocities of microtubules are independent of the local polarity and only depend
on the sliding velocity of the motors themselves.  This phenomenon is explained
by motors generating an active strain rate in regions of varying polarity. Thus,
microtubules can slide past each other without stressing the material since
$\nabla \mathbf v = -V\nabla\mathbf P$, see Eq.~{(\ref{eq:derived_stress})}.
Ultimately, the system consists of two inter-penetrating gels of microtubules of
opposite polarity that push off of each other, with each gel held together by
viscous coupling.  This is in contrast to dilute suspensions, in which the mass
fluxes of left moving and right moving microtubules locally balance, leading to
a strong dependence of sliding speed on local polarity.

Our theory predicts that microtubules in the network slide apart at the same
speed as isolated pairs of microtubules, independent of the local polarity, as
we observe in our experiments (Fig.~\ref{fig:experiment2}). This prediction does
not depend on any adjustable parameters and arises directly from the form of
Eqs.~(\ref{eq:force_velocity_relation},\ref{eq:MTForce}). In turn, the form of
Eq.~(\ref{eq:force_velocity_relation}) results from imposing that molecular
motors act uniformly along the length of microtubules. Our framework can be
extended to investigate microtubule networks crosslinked by motors with
different properties. For example, dynein accumulates at microtubule minus ends
and clusters those ends together \cite{tan2018cooperative}, leading to network
contractions \cite{foster2015active, foster2017connecting}. This can be
implemented in our theory by modifying Eq.~{(\ref{eq:force_velocity_relation})}
to include a preference for binding specific parts of the microtubule, and gives
rise to the $\mathcal Q$ and $\rho$ dependent active stresses that drive many
well studied phenomena in active matter systems.

Polarity independent sliding is also observed in spindles formed in {\it
Xenopus} egg extracts. These spindles consist of an array of microtubules which
are anti-parallel near the spindle center and highly polar at the spindle poles
\cite{brugues2012nucleation, yu2014measuring}. Microtubules in these spindles
continually slide toward spindle poles, a motion believed to be driven by the
molecular motor kinesin-5 \cite{mitchison2005mechanism}. The speed of
microtubule sliding is relatively constant throughout the spindle, particularly
so when dynein is inhibited \cite{burbank2007slide,yang2008regional}, and is
approximately equal to the speed that kinesin-5 slides apart pairs of
anti-parallel microtubules {\it in vitro } \cite{kapitein2005bipolar}. Our work
suggests that this phenomenon naturally arises if microtubules in the spindle
are heavily crosslinked and if kinesin-5 acts uniformly along the length of
microtubules. We speculate that spindles are self-straining like the
XCTK2-microtubule gel presented here.

Many cytoskeletal networks are heavily crosslinked. For instance, the
contractility of actin networks has been explained as emerging from its heavily
crosslinked nature \cite{ronceray2016fiber, belmonte2017theory}. The work
presented here is an important step towards predicting the material properties
of actively crosslinked materials from the properties of their constituents and
will enable the design of actively crosslinked networks from first principles.
\\
\noindent{\bf Data availability:}
Figs.~(\ref{fig:experiment} and \ref{fig:experiment2}) are based on microscopy
    data. The raw data are available from the authors upon reasonable request.
\\
\noindent{\bf Author Contributions:}
SF, MJS and DJN developed the theory.
BL, PJF, SEM, CEW and ZD performed experiments and provided materials.
SF, BL, DJN, and MJS wrote the paper with input from all authors.
\\
\noindent{\bf Acknowledgements:}
CEW acknowledges support by NIH R35GM122482.
DJN acknowledges the Kavli Institute for Bionano Science and Technology at
Harvard University, and National Science Foundation grants PHY-1305254,
PHY-0847188, DMR-0820484, and DBI-0959721. PJF acknowledges the
Gordon and Betty Moore Foundation for support as a Physics of Living Systems
Fellow through Grant GBMF4513. MJS acknowledges National Science
Foundation Grants DMR-0820341 (NYU MRSEC), DMS-1463962, and DMS-1620331.
We acknowledge support from the NSF MRSEC DMR-1420382.

\newpage

\begin{thebibliography}{1}

\bibitem{walczak1997xctk2}
Claire~E Walczak, Suzie Verma, and Timothy~J Mitchison.
\newblock Xctk2: a kinesin-related protein that promotes mitotic spindle
  assembly in xenopus laevis egg extracts.
\newblock {\em The Journal of Cell Biology}, 136(4):859--870, 1997.

\bibitem{ems2013aurora}
Stephanie~C Ems-McClung, Sarah~G Hainline, Jenna Devare, Hailing Zong, Shang
  Cai, Stephanie~K Carnes, Sidney~L Shaw, and Claire~E Walczak.
\newblock Aurora b inhibits mcak activity through a phosphoconformational
  switch that reduces microtubule association.
\newblock {\em Current Biology}, 23(24):2491--2499, 2013.

\bibitem{hentrich2010microtubule}
Christian Hentrich and Thomas Surrey.
\newblock Microtubule organization by the antagonistic mitotic motors kinesin-5
  and kinesin-14.
\newblock {\em The Journal of Cell Biology}, 189(3):465--480, 2010.

\bibitem{foster2017connecting}
Peter~J Foster, Wen Yan, Sebastian F{\"u}rthauer, M~Shelley, and Daniel~J
  Needleman.
\newblock Connecting macroscopic dynamics with microscopic properties in active
  microtubule network contraction.
\newblock {\em New Journal of Physics}, 19, 2017.

\end{thebibliography}


\begin{thebibliography}{10}

\bibitem{marchetti2013hydrodynamics}
MC~Marchetti, JF~Joanny, S~Ramaswamy, TB~Liverpool, J~Prost, Madan Rao, and
  R~Aditi Simha.
\newblock Hydrodynamics of soft active matter.
\newblock {\em Reviews of Modern Physics}, 85(3):1143, 2013.

\bibitem{needleman2017active}
Daniel Needleman and Zvonimir Dogic.
\newblock Active matter at the interface between materials science and cell
  biology.
\newblock {\em Nature Reviews Materials}, 2(9):17048, 2017.

\bibitem{alberts_2002_book}
B.~Alberts, D.~Bray, J.~Lewis, M.~Raff, K.~Roberts, and J.D. Watson.
\newblock {\em {Molecular Biology of the Cell}}.
\newblock Garland, 4th edition, 2002.

\bibitem{kruse2005generic}
Karsten Kruse, Jean-Francois Joanny, Frank J{\"u}licher, Jacques Prost, and Ken
  Sekimoto.
\newblock Generic theory of active polar gels: a paradigm for cytoskeletal
  dynamics.
\newblock {\em The European Physical Journal E}, 16(1):5--16, 2005.

\bibitem{joanny2007hydrodynamic}
JF~Joanny, F~J{\"u}licher, K~Kruse, and J~Prost.
\newblock Hydrodynamic theory for multi-component active polar gels.
\newblock {\em New Journal of Physics}, 9(11):422, 2007.

\bibitem{julicher2018hydrodynamic}
Frank J{\"u}licher, Stephan~W Grill, and Guillaume Salbreux.
\newblock Hydrodynamic theory of active matter.
\newblock {\em Reports on Progress in Physics}, 81(7), 2018.

\bibitem{mitchison2005mechanism}
TJ~Mitchison.
\newblock Mechanism and function of poleward flux in Xenopus extract meiotic
  spindles.
\newblock {\em Philosophical Transactions of the Royal Society of London B:
  Biological Sciences}, 360(1455):623--629, 2005.

\bibitem{burbank2007slide}
Kendra~S Burbank, Timothy~J Mitchison, and Daniel~S Fisher.
\newblock Slide-and-cluster models for spindle assembly.
\newblock {\em Current Biology}, 17(16):1373--1383, 2007.

\bibitem{yang2008regional}
Ge~Yang, Lisa~A Cameron, Paul~S Maddox, Edward~D Salmon, and Gaudenz Danuser.
\newblock Regional variation of microtubule flux reveals microtubule
  organization in the metaphase meiotic spindle.
\newblock {\em The Journal of Cell Biology}, 182(4):631--639, 2008.

\bibitem{naganathan2018morphogenetic}
Sundar~Ram Naganathan, Sebastian F\"urthauer, J.~Rodriguez, B.~T. Fievet, Frank
  J\"ulicher, J.~Ahringer, C.~V. Cannistraci, and S.W. Grill.
\newblock Morphogenetic degeneracies in the actomyosin cortex.
\newblock {\em Elife}, 7:e37677, 2018.

\bibitem{roostalu2018determinants}
Johanna Roostalu, Jamie Rickman, Claire Thomas, Fran{\c{c}}ois N{\'e}d{\'e}lec,
  and Thomas Surrey.
\newblock Determinants of polar versus nematic organization in networks of
  dynamic microtubules and mitotic motors.
\newblock {\em Cell}, 175(3):796--808, 2018.

\bibitem{bray2001cell}
Dennis Bray.
\newblock Cell movements: from molecules to motility, 2001.

\bibitem{furthauer2012active}
S~F{\"u}rthauer, M~Strempel, SW~Grill, and F~J{\"u}licher.
\newblock Active chiral fluids.
\newblock {\em The European Physical Journal. E, Soft matter}, 35:89, 2012.

\bibitem{thampi2013velocity}
Sumesh~P Thampi, Ramin Golestanian, and Julia~M Yeomans.
\newblock Velocity correlations in an active nematic.
\newblock {\em Physical Review Letters}, 111(11):118101, 2013.

\bibitem{salbreux2009hydrodynamics}
Guillaume Salbreux, Jacques Prost, and Jean-Francois Joanny.
\newblock Hydrodynamics of cellular cortical flows and the formation of
  contractile rings.
\newblock {\em Physical Review Letters}, 103(5):058102, 2009.

\bibitem{mayer2010anisotropies}
Mirjam Mayer, Martin Depken, Justin~S Bois, Frank J{\"u}licher, and Stephan~W
  Grill.
\newblock Anisotropies in cortical tension reveal the physical basis of
  polarizing cortical flows.
\newblock {\em Nature}, 467(7315):617--621, 2010.

\bibitem{naganathan2014active}
Sundar~Ram Naganathan, Sebastian F{\"u}rthauer, Masatoshi Nishikawa, Frank
  J{\"u}licher, and Stephan~W Grill.
\newblock Active torque generation by the actomyosin cell cortex drives
  left--right symmetry breaking.
\newblock {\em Elife}, 3:e04165, 2014.

\bibitem{brugues2014physical}
Jan Brugu{\'e}s and Daniel Needleman.
\newblock Physical basis of spindle self-organization.
\newblock {\em Proceedings of the National Academy of Sciences},
  111(52):18496--18500, 2014.

\bibitem{kruse2000actively}
Karsten Kruse and F~J{\"u}licher.
\newblock Actively contracting bundles of polar filaments.
\newblock {\em Physical Review Letters}, 85(8):1778, 2000.

\bibitem{aranson2005pattern}
Igor~S Aranson and Lev~S Tsimring.
\newblock Pattern formation of microtubules and motors: Inelastic interaction
  of polar rods.
\newblock {\em Physical Review E}, 71(5):050901, 2005.

\bibitem{liverpool2005bridging}
Tanniemola~B Liverpool and M~Cristina Marchetti.
\newblock Bridging the microscopic and the hydrodynamic in active filament
  solutions.
\newblock {\em EPL (Europhysics Letters)}, 69(5):846, 2005.

\bibitem{liverpool2008hydrodynamics}
Tanniemola~B Liverpool and M~Cristina Marchetti.
\newblock Hydrodynamics and rheology of active polar filaments.
\newblock In {\em Cell Motility}, pages 177--206. Springer, 2008.

\bibitem{saintillan2008instabilitiesPRL}
David Saintillan and Michael~J Shelley.
\newblock Instabilities and pattern formation in active particle suspensions:
  kinetic theory and continuum simulations.
\newblock {\em Physical Review Letters}, 100(17):178103, 2008.

\bibitem{saintillan2008instabilitiesrPHYSFLUID}
David Saintillan and Michael~J Shelley.
\newblock Instabilities, pattern formation, and mixing in active suspensions.
\newblock {\em Physics of Fluids}, 20(12):123304, 2008.

\bibitem{saintillan2013active}
David Saintillan and Michael~J Shelley.
\newblock Active suspensions and their nonlinear models.
\newblock {\em Comptes Rendus Physique}, 14(6):497--517, 2013.

\bibitem{foster2015active}
Peter~J Foster, Sebastian F{\"u}rthauer, Michael~J Shelley, and Daniel~J
  Needleman.
\newblock Active contraction of microtubule networks.
\newblock {\em eLife}, page e10837, 2015.

\bibitem{gao2015multiscalepolar}
Tong Gao, Robert Blackwell, Matthew~A. Glaser, M.~D. Betterton, and Michael~J.
  Shelley.
\newblock Multiscale polar theory of microtubule and motor-protein assemblies.
\newblock {\em Phys. Rev. Lett.}, 114:048101, Jan 2015.

\bibitem{heidenreich2016hydrodynamic}
Sebastian Heidenreich, J{\"o}rn Dunkel, Sabine~HL Klapp, and Markus B{\"a}r.
\newblock Hydrodynamic length-scale selection in microswimmer suspensions.
\newblock {\em Physical Review E}, 94(2):020601, 2016.

\bibitem{maryshev2018kinetic}
Ivan Maryshev, Davide Marenduzzo, Andrew~B Goryachev, and Alexander Morozov.
\newblock Kinetic theory of pattern formation in mixtures of microtubules and
  molecular motors.
\newblock {\em Physical Review E}, 97(2):022412, 2018.

\bibitem{belmonte2017theory}
Julio~M Belmonte, Maria Leptin, and Fran{\c{c}}ois N{\'e}d{\'e}lec.
\newblock A theory that predicts behaviors of disordered cytoskeletal networks.
\newblock {\em Molecular Systems Biology}, 13(9):941, 2017.

\bibitem{hentrich2010microtubule}
Christian Hentrich and Thomas Surrey.
\newblock Microtubule organization by the antagonistic mitotic motors kinesin-5
  and kinesin-14.
\newblock {\em The Journal of Cell Biology}, 189(3):465--480, 2010.

\bibitem{foster2017connecting}
Peter~J Foster, Wen Yan, Sebastian F{\"u}rthauer, M~Shelley, and Daniel~J
  Needleman.
\newblock Connecting macroscopic dynamics with microscopic properties in active
  microtubule network contraction.
  \newblock {\em New Journal of Physics}, 19(12):125011, 2017.

\bibitem{tan2018cooperative}
Ruensern Tan, Peter~J Foster, Daniel~J Needleman, and Richard~J McKenney.
\newblock Cooperative accumulation of dynein-dynactin at microtubule minus-ends
  drives microtubule network reorganization.
\newblock {\em Developmental Cell}, 44(2):233--247, 2018.

\bibitem{brugues2012nucleation}
Jan Brugu{\'e}s, Valeria Nuzzo, Eric Mazur, and Daniel~J Needleman.
\newblock Nucleation and transport organize microtubules in metaphase spindles.
\newblock {\em Cell}, 149(3):554--564, 2012.

\bibitem{yu2014measuring}
Che-Hang Yu, Noah Langowitz, Hai-Yin Wu, Reza Farhadifar, Jan Brugues, Tae~Yeon
  Yoo, and Daniel Needleman.
\newblock Measuring microtubule polarity in spindles with second-harmonic
  generation.
\newblock {\em Biophysical Journal}, 106(8):1578--1587, 2014.

\bibitem{kapitein2005bipolar}
Lukas~C Kapitein, Erwin~JG Peterman, Benjamin~H Kwok, Jeffrey~H Kim, Tarun~M
  Kapoor, and Christoph~F Schmidt.
\newblock The bipolar mitotic kinesin eg5 moves on both microtubules that it
  crosslinks.
\newblock {\em Nature}, 435(7038):114, 2005.

\bibitem{ronceray2016fiber}
Pierre Ronceray, Chase~P Broedersz, and Martin Lenz.
\newblock Fiber networks amplify active stress.
\newblock {\em Proceedings of the National Academy of Sciences},
  113(11):2827--2832, 2016.

\end{thebibliography}
\end{document}


\title{Methods}
\maketitle
\renewcommand{\theequation}{S.\Roman{section}-\arabic{equation}}
\renewcommand{\bibnumfmt}[1]{[S#1]}
\renewcommand{\citenumfont}[1]{S#1}

\section{Expansion of the microtubule Force balance}
\noindent
In the following, we illustrate in more detail how Eq.~{(\ref{eq:MTForce})} is
expanded to obtain Eq.~{(\ref{eq:MTiforce1})}. We start by rewriting
Eq.~{(\ref{eq:MTForce})} in terms of variables rescaled to the system size $\ell$,
and inserting the force velocity relation
Eq.~{(\ref{eq:force_velocity_relation})}.
\begin{eqnarray}
    \mathbf F_i &=&
    -\gamma\ell^6 \int\limits_{\hat L/2}^{-\hat L/2} d\hat s_i
    \sum\limits_j
    \int\limits_{\hat L/2}^{-\hat L/2} d\hat s_j
    \int d\mathbf p \int\limits_{\hat \Omega(\ell(\mathbf{\hat x_i} +\hat s_i
    \mathbf p_i))}d\hat x^3
    \left\{\delta(\ell(\mathbf{\hat x} -\mathbf{\hat x_j}-\hat s_j\mathbf{p_j})\hat{\mathbf{F}}_{ij}
\right\},
    \label{eq:MTiforcenondim}
\end{eqnarray}
where we introduced the length rescaled variables $\ell \hat x = x$. Most
importantly the rescaled microtubule length $\hat L = L/\ell$ and the
force $\hat{\mathbf{F}}_{ij} = \mathbf{F}_{ij}/\ell$.
We next expand the argument of the integral by writing
\begin{eqnarray}
    \delta(\ell (\hat{\mathbf{x}}-\hat{\mathbf{x}}_j-\hat s_j \mathbf
    p))\hat{\mathbf F}_{ij} = \delta(\ell (\hat{\mathbf{x}}-\hat{\mathbf{x}}_j))\hat{\mathbf F}_{ij}
    - \ell\hat s_j \mathbf p_j \cdot\nabla \delta(\ell
    (\hat{\mathbf{x}}-\hat{\mathbf{x}}_j))\hat{\mathbf
    F}_{ij}+\mathbf{O}(\varepsilon^3)
    \label{eq:expansion1rule}
\end{eqnarray}
where we use that $\hat s_j$ and $\hat{\mathbf F}_{ij}$ are small quantities of
order $\varepsilon$. Then Eq.~\ref{eq:expansion1rule} becomes,
\begin{eqnarray}
    \mathbf F_i &=&
    -\gamma\ell^6 \int\limits_{\hat L/2}^{-\hat L/2} d\hat s_i
    \sum\limits_j
    \int\limits_{\hat L/2}^{-\hat L/2} d\hat s_j
    \int d\mathbf p \int\limits_{\hat \Omega(\ell(\mathbf{\hat x_i} +\hat s_i
    \mathbf p_i))}d\hat x^3
    \left\{\delta(\ell(\mathbf{\hat x} -\mathbf{\hat
            x_j})\left(\hat{\mathbf{F}}_{ij}-\ell\hat s_j \nabla\cdot\mathbf
            p_j\hat{\mathbf{F}}_{ij}\right)
        \right\}\nonumber\\ &&+ \mathcal{O}(\varepsilon^8),
    \label{eq:MTiforcenondi2}
\end{eqnarray}
where we used that $\int\limits_{\hat\Omega} d\hat x^3 \simeq \varepsilon^3$.
We next use that
\begin{eqnarray}
    \int\limits_{\hat\Omega(\ell(\hat{\mathbf{x}}_i +\hat s_i \mathbf{p_i}))}
    d\hat x^3 X =  
    \int\limits_{\hat\Omega(\ell(\hat{\mathbf{x}}))}
    d\hat x^3 \left(X + \ell\nabla\cdot \hat s_i \mathbf{p_i} X \right) +
    \mathcal{O}(\varepsilon^5 X)\nonumber\\
    \label{eq:expansionrule2}
\end{eqnarray}
and finally find
\begin{eqnarray}
    \mathbf F_i &=&
    -\gamma\ell^6 \int\limits_{\hat L/2}^{-\hat L/2} d\hat s_i
    \sum\limits_j
    \int\limits_{\hat L/2}^{-\hat L/2} d\hat s_j
    \int d\mathbf p \int\limits_{\hat \Omega(\ell(\mathbf{\hat x_i}))}d\hat x^3
    \left\{\delta(\ell(\mathbf{\hat x} -\mathbf{\hat
            x_j})\left(\hat{\mathbf{F}}_{ij}-\ell \nabla\cdot\left(\hat
            s_j\mathbf p_j -\hat s_i\mathbf p_i\right)\hat{\mathbf{F}}_{ij}\right)
        \right\}\nonumber\\ &&+ \mathcal{O}(\varepsilon^8),
    \label{eq:MTiforcenondi3}
\end{eqnarray}
executing the integrals over $\hat s_i, \hat s_j$ and returning to natural
variables yields Eq.~{(\ref{eq:MTiforce1})}.

\section{Torque Balance}
\noindent
In this section we derive equation (\ref{eq:rotation_rates})
from the total torque
\begin{eqnarray}
    \mathbf T_i = \int\limits_{-L/2}^{L/2} ds_i
    \sum\limits_j
    \int\limits_{-L/2}^{L/2} ds_j 
    \int d\mathbf p
    \int\limits_{\Omega(\mathbf x_i + s_i \mathbf p_i )}dx^3
    \left\{
\delta(\mathbf x -\mathbf x_j-s_j \mathbf p_j)\delta(\mathbf p -\mathbf p_j)
\mathbf T_{ij}\right\}=0,
    \label{eq:MTTorque}
\end{eqnarray}
on the $i$-th microtubule. We start by rewriting (\ref{eq:MTTorque}) using
$\mathbf\Gamma_{ij} = \nu \mathbf p_i\times\mathbf p_j (\mathbf p_i\cdot\mathbf p_j)$ and the expression for $\mathbf F_{ij}$ from
Eq.~{(\ref{eq:force_velocity_relation})},
\begin{eqnarray}
    \mathbf T^{(0)}_i &=& 
    -\gamma \mathbf p_i \times
    \int\limits_{-L/2}^{L/2}ds_i
    \sum\limits_j
    \int\limits_{-L/2}^{L/2}ds_j\int
    d\mathbf p\int\limits_{\Omega(\mathbf x_i +s_i\mathbf
    p_i)}dx^3
    \nonumber\\ &&
    \left\{ s_i  
    \delta(\mathbf x- \mathbf x_j -s_j \mathbf p_j)\delta(\mathbf p-\mathbf
    p_j)\right.
    \left. \left(\mathbf v_i + s_i\mathbf p_i -\mathbf v_j - s_j\mathbf p_j -V \mathbf
p_j\right)\right\}
    \label{eq:MTTorque1}
\end{eqnarray}
where we used $\mathbf p_i \times \mathbf p_i =0$ and
\begin{eqnarray}
    \mathbf T^{(1)}_i &=& 
    \int\limits_{-L/2}^{L/2}ds_i\int\limits_{-L/2}^{L/2}ds_j\int
    d\mathbf p\int\limits_{\Omega(\mathbf x_i +s_i\mathbf
    p_i)}dx^3\left\{\mathbf p_i\times\mathbf p_j(\mathbf p_i\cdot\mathbf
    p_j)\right\}.
    \label{eq:MTTorque1a}
\end{eqnarray}
For notational convenience we separated  the total MT torque
$\mathbf T_i = \mathbf T_i^{(0)} + \mathbf T_i^{(1)}$ into parts generated
by crosslinking forces $\mathbf T_i^{(0)}$ and by crosslinking torques $\mathbf
T_i^{(1)}$. From here we follow the expansion steps outlined for the force
balance in the main text and obtain
\begin{eqnarray}
    \mathbf T^{(0)}_i &=& 
    -\gamma\frac{L^4}{12} \int\limits_{\Omega(\mathbf x_i)}dx^3 \rho(\mathbf x) \mathbf p_i
    \times\dot{\mathbf p}_i
    -\gamma\frac{L^4}{12} (\mathbf p_i\cdot\mathbf\nabla) \mathbf p_i \times \int\limits_{\Omega(\mathbf x_i)}dx^3
    \rho(\mathbf x)\left(\mathbf v_i  -\mathbf v(\mathbf x) -V \mathbf
    P(\mathbf x)\right).
    \label{eq:MTTorque2}
\end{eqnarray}
and 
\begin{eqnarray}
    \mathbf T^{(1)}_i &=& \rho \hat \nu \mathbf p_i\times\left(\mathbf p_i\cdot
    \mathcal{Q}\right),
    \label{eq:MTTorque2a}
\end{eqnarray}
with $\hat \nu = \nu L^2\frac{4\pi R^3}{3}$.
To simplify this expression consider
\begin{eqnarray}
    &&\mathbf p_i \times [\mathbf v_i - \mathbf v(\mathbf x) - V P(\mathbf x)]\nonumber\\
    &=& \mathbf p_i \times [\mathbf v_i - \mathbf v(\mathbf x_i) +V( \mathbf p_i
        - \mathbf P(\mathbf x_i))+ \left(
            \mathbf v(\mathbf x_i)-\mathbf v(\mathbf x) +V(
    P(\mathbf x_i) - P(\mathbf x))\right)]\nonumber\\
    &=& \mathbf p_i \times
    [-\frac{L^2}{12}\frac{\nabla\rho}{\rho}\cdot(\mathbf p_i \dot{\mathbf p}_i
        -\mathcal{H})+ \left(
            \mathbf v(\mathbf x_i)-\mathbf v(\mathbf x) +V(
    P(\mathbf x_i) - P(\mathbf x))\right)],\nonumber\\
    \label{eq:help1}
\end{eqnarray}
where we used Eq.{(\ref{eq:final_flux})}. After inserting Eq.~{(\ref{eq:help1})} back
into Eq.~{(\ref{eq:MTTorque2})} and using the force balance equation
(\ref{eq:derived_stress}) we finally obtain
\begin{eqnarray}
    &&\alpha\rho\left(1-\frac{L^2}{12}\left(\mathbf p_i\cdot\mathbf\nabla\frac{\mathbf
    p_i\cdot\mathbf\nabla\rho}{\rho}\right)\right)\mathbf p_i\times\dot{\mathbf p}_i
    = \rho \hat \nu \mathbf p_i\times\left(\mathbf p_i\cdot \mathcal{Q}\right) +
    \alpha\frac{L^2}{12}(\mathbf p_i\cdot\mathbf\nabla)\mathbf
    p_i\times\mathbf\nabla\cdot(\rho \mathcal{H}),
    \label{eq:TB_full}
\end{eqnarray}
which becomes Eq.~{(\ref{eq:rotation_rates})} after dropping higher order terms.

\section{Experimental Procedure}
To study the behavior of highly crosslinked active gels we performed
experiments on an {\it in vitro} system of purified components. We
made solutions of tubulin and mCitrine-XCTK2, a Kinesin-14 molecular motor
capable of cross-linking and sliding aligned microtubules. To image the tubulin,
it was fluorescently labeled with 0.7\% of Atto 565 NHS-Ester. We then added
paclitaxel, which nucleates and stabilizes microtubules; and rapidly loaded the
sample into a thin rectangular microfluidic chamber with approximate dimensions
of $0.1\times1\times20\mathrm{mm}^3$ . The microtubule-motor mixture
spontaneously self-organized into a highly aligned
macroscopic gel, oriented along the long axis of the microfluidic
channel (Fig.~\ref{fig:experiment} A).  

We photo-bleached fluorescent microtubules with lines perpendicular to the axis
of gel alignment (see Imaging and Bleaching section of SI). Photobleaching
bleached the Atto 565 labeled microtubules, but did not change the structure of
the co-located mCitrine-XCTK2 as seen in Supplementary (Fig.~\ref{fig:Sup1}),
demonstrating that the photobleaching did not ablate the microtubules. As
bleaching marks a subset of microtubules, it allows their subsequent motions to
be monitored. We observe that bleached lines split into two, and this subsequent
pair of lines moved apart along the direction of nematic alignment
(Fig.~\ref{fig:experiment} B, left), indicating that microtubules in the gel are
continually sliding relative to each other. 

\begin{figure}[h] \centering
	\includegraphics[width=0.4\columnwidth]{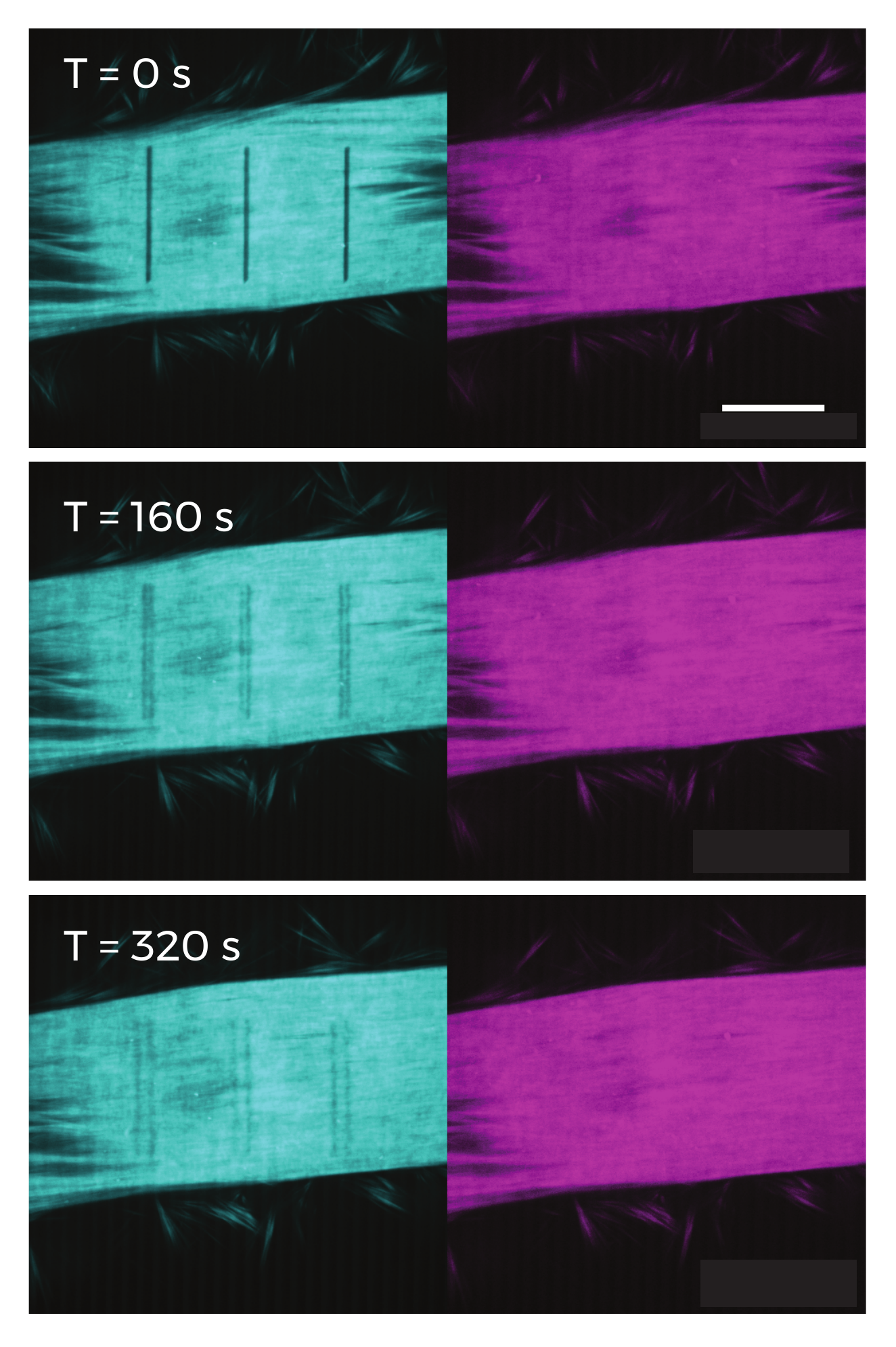} 
        \caption{ Photo-bleached lines label the Atto 565 labeled microtubules
    (Cyan, Left), but do not change the structure of the co-located
    mCitrine-XCTK2 (Magenta, Right). Scale bar is 100 $\mu$m.  }
\label{fig:Sup1} \end{figure}

\section{Cloning, expression and purification of mCit-XCTK2}
Full-length amino-terminally tagged mCitrine XCTK2 (mCit-XCTK2) was expressed in
and purified from baculovirus infected Sf-9 cells (Invitrogen) using
conventional chromatography as previously described\cite{walczak1997xctk2}. Baculovirus stocks were
created from pFB-mCit-XCTK2 using the Bac-to-Bac® System (Invitrogen).
pFB-mCit-XCTK2 was created from pFB-mCit-DEST\cite{ems2013aurora} and pENTR-XCTK2 using the
Gateway® System. The full-length coding sequence of XCTK2 was amplified by PCR
and inserted into pENTR using the pENTR™/D-TOPO kit (Invitrogen) to create
pENTR-XCTK2. The plasmids are available from the authors upon reasonable request.
\section{Chamber Fabrication}
\noindent
Imaging chambers consisting of a slide and coverslip with a $0.9$mm channel in a
110$\mu$m PDMS spacer. Briefly, masters were made by first covering a slide with
polyester film (Schein Holographics) to prevent PDMS adhesion. A $0.9\mathrm{mm}
\times 47.5\mathrm{mm}$ rectangle was cut from adhesive vinyl (3M Gerber InstaChange) using a
die cutter (Silhouette Cameo), and adhered to the polyester coated slide. PDMS
(Dow Corning, 1:10 mixing ratio) was poured over the master and degassed before
a second, uncoated, slide was placed on top and secured in place using binder
clips. Sandwiched slides were baked overnight at $60^\circ$C. The slides were then
separated, leaving a slide with adhered PDMS and the polyester coated slide,
which was discarded. The PDMS coated slide and an
$18\mathrm{mm}\times18\mathrm{mm}$ coverslip were each
treated with air plasma for 1 minute using a corona device before bonding.
Chambers were loaded with passivation mix consisting of 5mg/mL BSA (JT Baker)
and 5\% (w/v) Pluronic F-127 (Sigma P2443) and incubated overnight at $8^\circ$C.
Before use, passivation mix was flushed from the chambers using BRB80.

\section{Self-organization Assay}
\noindent
Self-organization buffer (SOB) was adapted from (Hentrich 2010) \cite{hentrich2010microtubule} and prepared as
a 2x stock containing 40 mM PIPES (Sigma P6757), 2mM EGTA (Sigma E4378), 14.5mM
MgCl2 (Sigma M8266), 10mM ATP (Sigma A2383), 3mM GTP (Sigma G8877), 2mM
$\beta$-mercaptoethanol (Bio-Rad \#161-0710), 100mM KCl (Sigma 60130), 400mM Sucrose
(Sigma 84097), 61.5mM Glucose (Sigma D9434),  pH 6.85. Reaction mix consisting
of 10 $\mu$L 2x SOB, 0.6$\mu$L Gloxy, 1 $\mu$L mCit-XCTK2 (10 $\mu$M stock), 0.7 $\mu$L PK/LDH (Sigma P0294), 0.5
$\mu$L PEP (MP Biomedicals, IC15187280), 4.15 $\mu$L Milli-Q H$_2$O, and 0.25$\mu$L Atto-565
labeled tubulin (20 $\mu$M stock) and 2.32 $\mu$L unlabeled tubulin (300$\mu$M stock).
Reactions were initiated by the addition of 0.4 $\mu$L paclitaxel (Sigma T7191, 100 $\mu$M
stock). Immediately after paclitaxel addition, the reaction mix was loaded into
a passivated imaging chamber, and the chamber inlet and outlet were sealed using
candle wax. 

We estimate the initial concentration of tubulin to be ~36 $\mu$M. We estimate
from confocal data with slices taken every 3$\mu$m that the material contracts
to ~24 $\mu$m in height and ~200 $\mu$m in width. We also estimate from taking
background florescence data that roughly 25\% of the tubulin remains outside the
contracting network. This back of the envelope calculation gives a tubulin
concentration in the final state to be 558 $\mu$M. Based off previous MT length
distribution measurements in similar experiments \cite{foster2017connecting}
suggesting that the microtubules will be very roughly ~6 $\mu$m , we can
estimate that there are ~20,000 tubulin units per microtubule. This gives a
number concentration of 27.9 nM microtubules, or $1.7 \times 10^{10}$
microtubules per $\mu$L. Given the 25 nm diameter of a microtubule and the
estimated length, this results in a volume fraction of 5\%. Since there are 17
microtubules per $\mu$m$^3$, we estimate that the average distance between
microtubules is 389 nm.

\section{Imaging and Bleaching}
\noindent
The imaging system consists of an inverted microscope (Eclipse Ti, Nikon), with
a femtosecond Ti:sapphire pulsed laser (Mai-Tai, Spectra-Physics) for excitation (800 nm
wavelength, 80 MHz repetition rate, ~70 fs pulse width), a commercial scanning
system (DCS-120, Becker \& Hickl), and hybrid detectors (HPM-100-40, Becker \&
Hickl). The excitation laser was collimated by a telescope to avoid any power
loss at the XY galvanometric mirror scanner and to fill the back aperture of a
water-immersion objective (CFI Apo 40 WI, NA 1.25, Nikon).

Two-photon fluorescence was imaged with an emission filter wheel controlled by
$\mu$Manager (Edelstein et al., 2014) that contained filters for mCitrine-labels on
XCTK2 (FF01-510/42-25, Semrock) and Atto565-labeled tubulin (FF01-607/36-25,
Semrock). The optical path contained a short-pass filter (FF01-650/SP-25,
Semrock) to block the fundamental laser wavelength. Samples were imaged at room
temperature ($\simeq 21^\circ$C).

During imaging signal was acquired for ~10 s of integration time at a spatial
resolution of 512x512 pixels over a 110x110 $\mu\mathrm{m}^2$ field of view. The maximum scan
rate of the DCS-120 at this resolution is ~2 frames/s. The power of the
excitation laser was adjusted to 10 mW at the objective.

To bleach, the excitation laser was adjusted to 40 mW as measured at the
objective. The spatial resolution was decreased to 16x16 pixels over a 440x440
$\mu\mathrm{m}^2$ field of view. A 30s scan at this resolution results in 16 distinct bleached
lines at the wider FoV, and 4 distinct lines in the smaller FoV used for
imaging.  

Subsequent experiments to investigate possible three-dimensional considerations
of the material was performed using the same laser attached to a Spinning Disk
confocal microscope. This data was acquired on an EMCCD camera of 512 × 512
pixels (ImageM enhanced C9100-13; Hamamatsu Photonics). The images were using
$\mu$Manager software (Vale laboratory at University of California, San
Francisco) and Z-stack images were acquired by high-speed piezo-Z scanner (P-725
PIFOC; Physik Instrumente).

\section{Data Analysis}
\noindent
Fluorescence image data was manually cropped to include just one splitting line.
A temporal series of cropped images was read into Matlab for analysis. Data was
binned in the direction parallel to the bleach lines and fitted by a function of
two Gaussian functions plus a linear slope. A linear fit of the temporal series
of the location of the Gaussian functions was used to calculate relative
velocities between the two Gaussian functions. The amplitudes of the Gaussian
functions, $A_1$ and $A_2$,  were used to calculate experimental polarity as
$P_{exp} =|\frac{ A_1 - A_2}{A_1	+ A_2}|$. This code has been made
publicly available at \url{https://github.com/bezlemma}

\bibliography{../bibliography/bibliography.bib}
\bibliographystyle{unsrt}